\documentclass[floatfix, aip, jap, superscriptaddress, reprint]{revtex4-1}

\usepackage{graphicx}               
\usepackage{bm}                     
\usepackage{amsfonts,amssymb}       
\usepackage{dcolumn}                
\usepackage{rotating}

\begin{document}

\title{
The role of resonant bonding in governing the thermal transport properties of
two-dimensional black phosphorus
}

\author{Guangzhao~Qin}
\email{qin.phys@gmail.com}
\affiliation{Institute of Mineral Engineering, Division of Materials Science and Engineering, Faculty of Georesources and Materials Engineering, RWTH Aachen University, Aachen 52064, Germany}
\author{Ming~Hu}
\affiliation{Institute of Mineral Engineering, Division of Materials Science and Engineering, Faculty of Georesources and Materials Engineering, RWTH Aachen University, Aachen 52064, Germany}
\affiliation{Aachen Institute for Advanced Study in Computational Engineering Science (AICES), RWTH Aachen University, Aachen 52062, Germany}

\date{\today}


\maketitle

%
The thermal transport properties is attracting ever much interest both for its
role and relevance to practical implications, such as electronic
cooling, thermoelectrics, phase change memories\cite{AppliedPhysicsReviews.2014.1.1.011305},
thermal devices (diodes, transistors, logic gates)\cite{Phys.Rev.Lett..2004.93.18.184301},
\emph{etc}. 
Phonons, the collective excitation of lattice dynamics, play dominant role in
the thermal transport in semiconductors 
\cite{NatMater.2011.10.8.569-581}.
Fundamental insight into lattice dynamics and phonon transport is critical to
the efficient manipulation of heat flow, which is one of the appealing
thermophysical problems with enormous practical implications.
Intrinsic phonon-phonon scattering governed by phonon anharmonicity is the key
factor limiting the heat transfer ability, which is characterized by lattice
thermal conductivity ($\kappa$).
The wealth of fascinating phenomena arising from the phonon transport in
low-dimensional open new door for the scientific understanding and technological
applications opportunities of graphene and related two-dimensional (2D)
materials.
Phosphorene, being the single-layer counterpart of the bulk black phosphorus
(BP), is an elemental 2D semiconductor with novel high carrier mobility proved
by experiments and intrinsically large fundamental direct band gap in the
electronic structure ($\sim1.5\,\mathrm{eV}$)\cite{ACSNano84033nn501226z}.
The distinctive properties of phosphorene give rise to great prospective for its
applications as the active layer in nano-/opto-electronic devices, such as
field-effect transistors and photo-transistors\cite{BP_NATURE}.
The rapidly growing applications of phosphorene in nano-/opto-electronics and
thermoelectrics call for fundamental understanding of the thermal transport
properties, which would be of great significance to the design and development
of high-performance phosphorene based nano-devices.

Herein, we provide insight into phonon transport properties of phosphorene with
the deep analysis of the electronic structure and lattice dynamics.
Based on the demonstration of resonant bonding formation in phosphorene, insight
into the thermal transport is achieved by discussing the role of resonant
bonding in driving long-range interactions and strong phonon anharmonicity.
All the relevant calculations are performed based on the density functional
theory (DFT) using the projector augmented wave (PAW) method as implemented in
the Vienna \emph{ab initio} simulation package
(\texttt{\textsc{vasp}})\cite{PhysRevB.54.11169}.


\begin{figure*}[tb]
    \centering
    \includegraphics[width=0.90\linewidth]{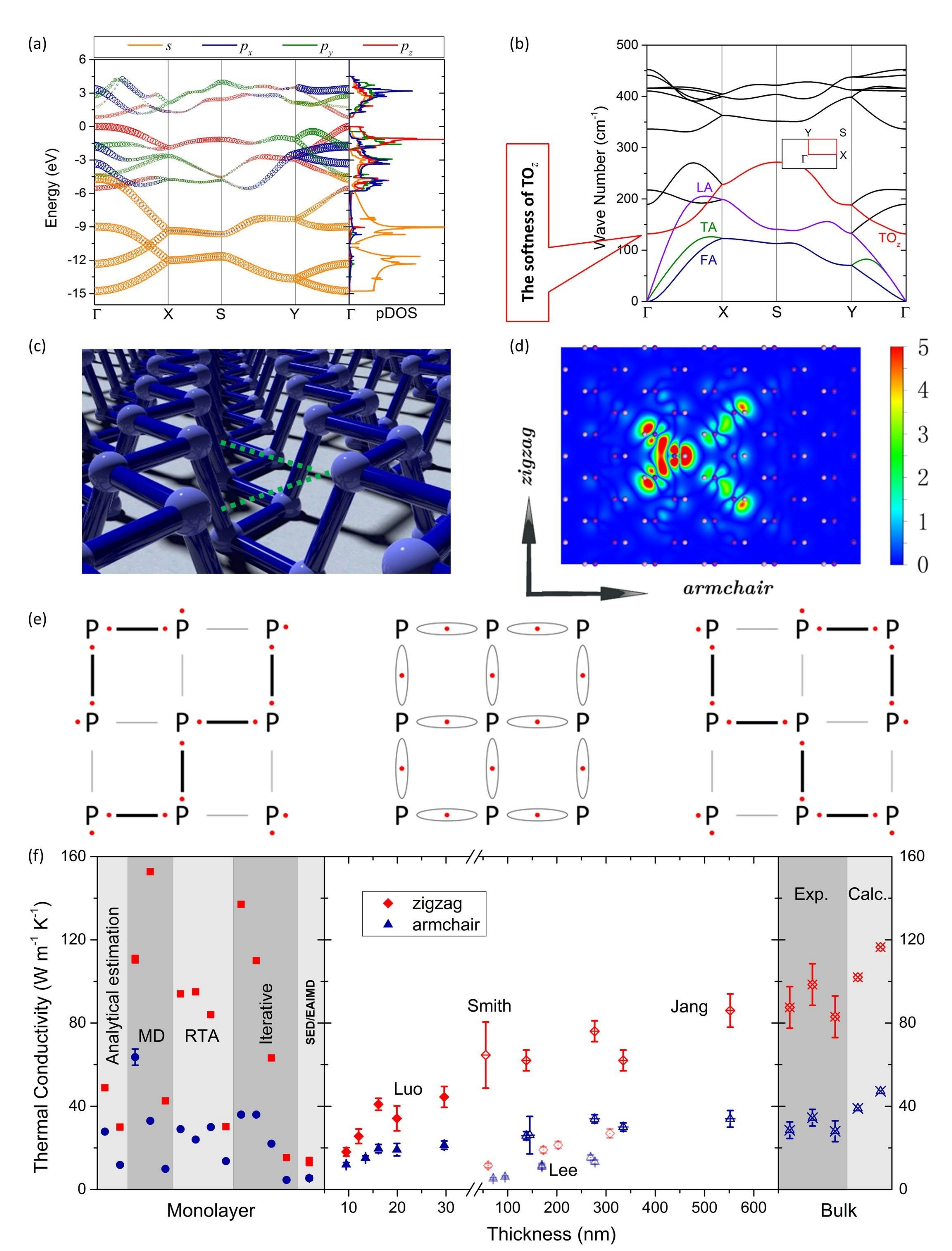}
\caption{\label{fig:resonant}
(a) Orbital projected electronic structure and density of states (pDOS) of phosphorene.
(b) Phonon dispersion with the soft TO$_z$ phonon branch highlighted.
(c) The perspective view of the geometry structure.
(d) The long-range interactions due to the resonant bonding as revealed by the
perturbed charge density by atomic displacement.
(e) The resonant bonding in an undistorted P phase is a hybrid form with
minimized energy (middle sub-figure) of different limiting cases for bonding
(left and right sub-figures).
(f) The comparison of thermal conductivity ($\kappa$) among the theoretical and
experimental results reported for monolayer, films with finite thickness, and
bulk forms.
}
\end{figure*}

The concept of resonance was introduced to achieve fundamental understanding on
certain properties of solids.
\cite{}
Generally, one can construct a many-body wave function as a linear combination
of different valence-bond configurations.
Thus, if $\phi_1$ and $\phi_2$ are two such valence-bond configurations, the
ground-state wave function $\psi$ can be expressed as
\begin{equation}
\psi = \frac{1}{(1+\alpha^2)^\frac{1}{2}} (\phi_1 + \alpha\phi_2)
\ ,
\end{equation}
where the mixing coefficient $\alpha$ is determined by optimizing the binding
energy or minimizing the total energy.
If $\alpha$ is very small or very large, the most stable ground state is
$\phi_1$ and $\phi_2$, respectively.
However, in the most interesting cases where $\alpha$ is of the order of unity,
the gound state involves both configurations and resonates between $\phi_1$ and
$\phi_2$.

Commonly, resonance occurs where electrons are insufficient to satisfy the
orbitals required for covalent bonding.
Resonant bonding is formed due to the resonant orbital occupations of electrons,
which is generally happened in group IV-VI compounds with rocksalt-like
structure\cite{NatMater.2008.7.12.972-977}.
Three valence $p$ electrons are available for each atom in average.
However, there are six nearest neighbours due to the rocksalt structure.
Thus, the bonding configuration based on the electron occupation is not unique.
The real bonding state is a hybridization among all the possible bonding
configurations of the three electrons forming six bonds.
We will show in the following how the resonant bonding is formed in phosphorene
and its effect on the lattice dynamics properties.

The typical premonitor for the formation of resonant bonding is the weak
$sp$-hybridization \cite{Phys.Rev.B.1973.8.2.660-667, NatMater.2008.7.8.653-658}.
To give a distinct view, the orbital-projected electronic structure and density
of states (pDOS) ($s$, $p_x$, $p_y$ and $p_z$) of phosphorene are plotted in
Fig.~\ref{fig:resonant}(a).
The main part of $s$-orbital is confined 9\,eV below the valence band maximum
(VBM), which hybridizes with $p_x/p_y/p_z$-orbital weakly.
The bonding states close to the VBM are dominated by the $p_z$-orbital, forming
a broad band with $\sim 6$\,eV, which hybridizes with the $p_x/p_y$-orbitals.
Due to the weak $sp$ hybridization in phosphorene [Fig.~\ref{fig:resonant}(a)],
only three $p$ electrons are available for the bonding between P atoms in
phosphorene.
Fig.~\ref{fig:resonant}(e) shows the (001) plane of phosphorus (P) in the ideal
rocksalt structure.
Two representative limiting configurations are shown on the left-/right-hand
sides in Fig.~\ref{fig:resonant}(e).
The superposition of all these configurations as schematically shown in the
middle of Fig.~\ref{fig:resonant}(e) is observed in reality.
The less valence electrons than the bonds results in the unsaturated covalent
bonding in phosphorene.
As a result, the single, half-filled $p$-band forms two bonds to the left and
right simultaneously (more than that allowed by the 8-$N$ rule), which is the so
called resonant bonding.
The real bonding state in phosphorene is a hybridization or resonance
among different electronic configurations of $p$-orbital occupations of
electrons.

In fact, the resonant bonding exists in a weakened form in phosphorene as
compared with the resonant bonding in regular rocksalt structures.
Phosphorene possesses the hinge-like structure, which is puckered as shown in
Fig.~\ref{fig:resonant}(c).
The puckered structure of phosphorene can be actually regarded as the deformed
rocksalt structure in two-dimensional\cite{Phys.Rev.B.2016.94.16.165445}, where
there are 5 bonds to be formed for each P atom.
Electron localization is increased by the distortion due to the no longer
energetically equivalence of the different limiting bonding configurations as
shown in Fig.~\ref{fig:resonant}(c).
Thus, the resonant bonding in phosphorene is weakened by the structural
distortion from perfect rocksalt structure of the hinge-like structure, which is
denoted to be Peierls-like distortion.
However, the weakened resonant bonding still has non-negligible impact on the
phonon transport properties of phosphorene.

One typical feature driven by the resonant bonding is the long-range
interactions.
The long-range interactions are evidently shown by the perturbation of charge
density distribution due to the atomic displacement
[Fig.~\ref{fig:resonant}(d)].
It shows that the disturbance can reach as far as $\sim$6\,\AA\ along the [110]
direction, corresponding to the collinear bonding direction of the resonant
bonding in the hinge-like structure of phosphorene.
The microscopic picture can be intuitively understood.
The resonant bonding is a superposition of different bonding configurations.
Due to the more bonding than that allowed by the 8-$N$ rule, the single,
half-filled $p$-band forms two bonds to the left and right simultaneously.
Consequently, one $p$-electron is shared by the two bonds, which leads to
long-range interactions.
For example, if one atom displaces along the $+x$ direction due to thermal
vibration, it perturbs the $p_x$ orbital of the adjacent atom.
Thus, the bonding electrons of the adjacent atom on the $-x$ side can easily
move to the $+x$ side since both sides are in the same $p_x$ orbital due to the
resonant bonding.
The orbital perturbation can persist over long range owing to the collinear
bonding characteristics and the large electronic polarizability 
\cite{AdvancedFunctionalMaterials.2011.21.12.2232-2239}.

Fig.~\ref{fig:resonant}(b) shows the phonon dispersion of phosphorene, where the
soft TO$_z$ phonon branch with frequency decreasing around the BZ center is
highlighted in red.
The softness of the TO$_z$ phonon branch is due to the long-range interactions
in phosphorene, which is caused by the resonant $p$ bonding as discussed above
and can be understood based on the 1D lattice chain model\cite{NatCommun.2014.5..3525}.
Previous studies demonstrated that the softening of TO phonon branch is
associated with strong phonon anharmonicity\cite{NatMater.2011.10.8.614-619}
which is measured by Gr\"{u}neisen parameter ($\gamma$).
It has been clearly revealed that the phonon anharmonicity of the soft TO$_z$ is
very strong, while those of LO$_y$ and TO$_x$ are
not\cite{Phys.Rev.B.2016.94.16.165445}.
In addition to single-layer phosphorene, the soft TO$_z$ phonon branch also
exist in few-layer phosphorene.

As discussed above, resonant bonding leads to long-range interactions, softened
TO$_z$ phonon branch and the corresponding strong phonon anharmonicity, which
have remarkable effect on the thermal transport properties of phosphorene.
For instance, due to the long-range interactions caused by the resonant
bonding, interactions up to $\sim 6$\,\AA\ are still fairly strong and a large
decrease of thermal conductivity of phosphorene at the cutoff distance of
$\sim$6\,\AA\ is observed.
Thus, the cutoff distance should be larger than 6\,\AA\ to get satisfactorily
converged results.
Another interesting point is the problem of convergence/divergence with sample
size ($Q$-grid).
A not large enough cutoff distance would yield a diverged thermal conductivity
\cite{NatMater.2011.10.8.569-581, PhysRevB.90.214302, Phys.Rev.B.2016.94.16.165445}.
If the interactions are truncated up to 4.4\,\AA\ in phosphorene, there exists a
size-dependence behavior for the thermal conductivity along the armchair
directions \cite{PhysRevB.90.214302}.
The reason for the size-dependent $\kappa$ of phosphorene primarily lies in the
quickly blowing up of the lifetime for phonons approaching the $\Gamma$ point of
the BZ.
The lifetime can be effectively suppressed when long-range interactions
($\sim$6\,\AA) caused by the resonant bonding are involved, and then the
$\kappa$ of phosphorene converges.

From the thermal conductivity of single-layer [left panel of
Fig.~\ref{fig:resonant}(f)], multi-layer phosphorene (middle panel), and bulk BP
(right panel), it can be clearly observed that the thermal conductivity of
phosphorene films starts from the high level of bulk BP and then decreases with
thickness decreasing.
It is anticipated that the thermal conductivity of single-layer phosphorene
should follow the trend and being lower than the experimental results for
phosphorene films with the minimized thickness of 9.5\,nm ($\sim$17-18 layers).
However, the results for single-layer phosphorene are only available from
theoretical reports, some of which are higher than the expectation and some are
lower.
Due to the limitations of the synthesis technique, the results for phosphorene
films with thickness smaller than 9.5\,nm are currently still unavailable.
The thermal conductivity of few-layer phosphorene with thinner thickness
approaching to the limit of single-layer is highly expected to obtain the
overall trend from bulk to single-layer and to verify the theoretical
predictions.

%
In summary, based on the analysis of electronic structure and lattice dynamics,
we provide fundamental insight into the thermal transport in phosphorene by
discussing the role of resonant bonding in driving long-range interactions and
strong phonon anharmonicity.
We reveal that the strong phonon anharmonicity is associated with the soft
transverse optical (TO) phonon modes and arises from the long-range interactions
driven by the orbital governed resonant bonding.
Our study highlights the physical origin of the phonon anharmonicity in
phosphorene, and also provides new insights into phonon transport from the view
of orbital states, which would be of great significance to the design and
development of high-performance phosphorene based nano-devices.

%

\bibliography{bibliography.bib}

\end{document}